%% file: RahbariScalo_WhitherTurbulence_2015.tex
\documentclass[graybox]{svmult}

\usepackage{graphicx,amssymb,amstext,amsmath}
\DeclareMathAlphabet{\mathcal}{OMS}{cmsy}{m}{n}
\usepackage[utf8]{inputenc} 
\usepackage{bm}
\usepackage{multicol} 
\usepackage{caption}
\usepackage{rotating}
\usepackage{cases}
\usepackage{mathptmx}       
\usepackage{helvet}         
\usepackage{courier}        
\usepackage{type1cm}        
\usepackage{makeidx}         
\usepackage{graphicx}        
\usepackage{multicol}        
\usepackage[bottom]{footmisc}
\DeclareSymbolFont{matha}{OML}{txmi}{m}{it}
\DeclareMathSymbol{\varv}{\mathord}{matha}{118}

\makeatletter
\newcommand*{\rom}[1]{\expandafter\@slowromancap\romannumeral #1@}
\makeatother

\usepackage{color}

\makeindex             

\begin{document}

\title*{Linear Stability Analysis of Compressible Channel Flow with Porous Walls}
\author{Iman Rahbari$^{*,1}$ and Carlo Scalo$^1$}
\institute{$(*)~$Corresponding author, \email{irahbari@purdue.edu}. $(1)~$School of Mechanical Engineering, Purdue University, West Lafayette, IN 47907-2045, USA.}
%
%
\maketitle

\abstract*{\input{Abstract.tex}
} 

\abstract{\input{Abstract.tex}}

\input{Introduction.tex}

\input{ProblemFormulation.tex}

\input{Results.tex}

\input{Conclusions.tex}

\begin{acknowledgement}
The authors would like to thank Prof. Johan Larsson for very kindly providing us with code, \emph{Hybrid} that was used to perform the high fidelity turbulent simulation. The authors have also significantly benefited from a generous computational allocation on Purdue's eighth on-campus supercomputer, \emph{Rice}.
\end{acknowledgement}
\input{referenc}
\end{document}

%% file: Abstract.tex

We have investigated the effects of permeable walls, modeled by linear acoustic impedance with zero reactance, on compressible channel flow via  linear stability analysis (LSA). Base flow profiles are taken from impermeable isothermal-wall laminar and turbulent channel flow simulations at bulk Reynolds number, $Re_b$= 6900 and Mach numbers, $M_b$ = 0.2, 0.5, 0.85. For a sufficiently high value of permeability, Two dominant modes are made unstable: a bulk pressure mode, causing symmetric expulsion and suction of mass from the porous walls (Mode 0); a standing-wave-like mode, with a pressure node at the centerline (Mode I). In the case of turbulent mean flow profiles, both modes generate additional Reynolds shear stresses augmenting the (base) turbulent ones, but concentrated in the viscous sublayer region; the trajectories of the two modes in the complex phase velocity space follow each other closely for values of wall permeability spanning two orders of magnitude, suggesting their coexistence. The transition from subcritical to supercritical permeability does not alter the structure of the two modes for the range of wavenumbers investigated, suggesting that wall permeability simply accentuates pre-existing otherwise stable modes. Results from the present investigation will inform the design of new compressible turbulent boundary layer control strategies via assigned wall-impedance.

%% file: Introduction.tex

\section{Introduction}
\label{sec:Intro}
In the present work we investigate the effects of porous walls modeled with a simple Darcy-like boundary condition derived as a particular case of linear acoustic impedance boundary condition (IBC) with zero reactance. The latter can be directly written in the time domain as
\begin{equation}
p' = R \, v'_n
\label{eq:purely_real_IBC}
\end{equation}
where $R$ is the impedance resistance, and $p'$ and $v'_n$ are the fluctuating pressure and wall-normal component of the velocity at the boundary (positive if directed away from the fluid side), normalized with the base density and speed of sound (unitary dimensionless base impedance). The present investigation is limited to a classic Linear Stability Analysis (LSA) approach and is inspired by the hydro-acoustic instability observed by Scalo, Bodart, and Lele \cite{ScaloBL_PoF_2015}, who performed numerical simulations of compressible channel flow with a three-parameter broadband wall-impedance \cite{TamA_AIAA_1996}. In the latter, the effective wall permeability is a broadband function of frequency tuned to allow maximum transpiration only at frequency of the large-scale turbulent energy containing eddies. Despite the lack of the frequency-dependency in \eqref{eq:purely_real_IBC}, the results shown in the present investigation provide a useful theoretical framework for the understanding of hydro-acoustic instabilities and a guide for future investigations in the manipulation of compressible boundary layer turbulence.

The effects of porous walls on shear flows has been a topic of formidable research effort, especially in the low-Mach-number limit. Lekoudis \cite{Lekoudis_AIAA_1978} performed LSA of an incompressible boundary layer over two types of permeable boundaries: one modeling a perforated surface over a large chamber (with stabilizing effects); the second one, modeling pores over independent cavities (with negligible effects). Jim\'enez et al. \cite{Jimenez2001Turbulent} performed channel flow simulations over active and passive porous walls observing the creation of large spanwise-coherent Kelvin-Helmholtz in the outer layer responsible for frictional drag increase; the near-wall turbulence production cycle was not found to be significantly altered \cite{jimenez1999autonomous}. More recently, Tilton and Cortelezzi \cite{Tilton_JFM_2008} investigated channel flow coated with finite-thickness homogeneous porous slabs. Two new unstable modes were discovered, one symmetric and one anti-symmetric, originating from the left branch of eigenvalue spectrum. Similar results have been found in the present investigation, which also supports the findings of Scalo \emph{et al.} \cite{ScaloBL_PoF_2015} who observed a complete reorganization of the near-wall turbulence, following the application of the tuned wall-impedance.

One of the earliest works in linear stability analysis of compressible flow has been carried out by Malik \cite{Malik_JCP_1990}, who investigated high-speed flat plate boundary layers up to freestream Mach numbers of 10. Duck et al \cite{Duck_JFM_1994} performed a similar investigation in the case of viscous compressible Couette flow, revealing fundamental differences with respect to unbounded cases. Hu and Zhong \cite{Hu_AIAA_1997,HuZhong_PoF_1998} investigated supersonic viscous Couette flow at finite Reynolds numbers finding two unstable acoustic inviscid modes when a region of locally supersonic flow, relative to the phase speed of the instability wave, is present. Their structure was deemed consistent with two unstable modes shown by Mack  \cite{Mack_TCFD_1990}, which are sustained by the acoustic interaction between the walls and the sonic line. 
Malik and co-workers  \cite{Malik_PR_2012} have also used modal and non-modal stability to study the effect of viscosity stratification on the stability of compressible Couette flow. Energy transfer from the mean flow to perturbations occurs at two locations: near the top wall and in the bulk of flow domain, associated with two different unstable modes. More recent efforts are focused on the transitional hypersonic boundary layer stability \cite{BresICF_JFM_2013,Wagner:2014PhD,WagnerHK_AIAA_2012,TullioS_PoF_2010,WartemannLS_AIAA_2012}, where porous walls are used as means to delay transition via acoustic energy absorption (allowed by a frequency-selective permeability)

The present work aims to inform future control strategies of fully-developed wall-bounded compressible boundary layer turbulence over porous walls. Comparison against previous LSA studies on channel and Couette flow is first discussed. Impermeable-wall laminar and turbulent channel flow data used as a base flow in the present study are then presented. Finally, the structure of the unstable modes triggered by the porous walls is discussed along with their predicted effect on the structure of near-wall turbulence.

%% file: ProblemFormulation.tex


\section{Problem Formulation}

The full nonlinear governing equations are first introduced, followed by their linearized counterpart, used for the present LSA. All quantities reported hereafter, including results from previous works are normalized with the channel's half-width (or total height in the case of Couette flow), the bulk density (constant for channel flow), and the speed of sound, temperature and dynamic viscosity at the wall.

\subsection{Governing Equations}

The dimensionless governing equations for conservation of mass, momentum and total energy are, respectively,

\begin{align}
&\frac{\partial}{\partial t} \left( \rho \right) +\frac{\partial}{\partial {x}_{j}} \left( \rho {u}_{j} \right)=0 \label{eq:NS_continuity}\\
&\frac{\partial}{\partial t} \left( \rho {u}_{i} \right) +\frac{\partial}{\partial {x}_{j}} \left( \rho {u}_{i} {u}_{j} \right)=-\frac{\partial}{\partial x_{i}} p + \frac{1}{\mathrm{Re}_a} \frac{\partial}{\partial x_{i}} \left( {\tau}_{ij} \right) + f_{1} \delta_{1i} \label{eq:NS_momentum} \\
&\frac{\partial}{\partial t} \left( \rho E \right) +\frac{\partial}{\partial {x}_{j}} \left[{u}_{j} \left( \rho E + p \right)  \right]=\frac{1}{\mathrm{Re}_a} \frac{\partial}{\partial x_{j}} \left( u_{i} {\tau}_{ij}+ {q}_{j} \right) + {f}_{1} u_{1} \label{eq:NS_energy}
\end{align}
where $x_1$, $x_2$ and $x_3$ (alternatively $x$, $y$, and $z$) are, respectively, the streamwise, wall-normal and spanwise coordinates, $u_i$ the velocity components in those directions. Thermodynamic pressure, density and temperature are related by the equation of state $p=\gamma^{-1} \rho T $, where $\gamma$ is the ratio of specific heats. $E$ is the total energy per unit mass and $Re_a$ the Reynolds number based on the reference length, (either the channel's half-width or wall-to-wall distance for Couette flow), and speed of sound at the wall temperature, which is related to the bulk Mach and Reynolds numbers via $Re_b = M_b\,Re_a$. The viscous stress tensor and conductive heat fluxes are
\begin{eqnarray}
\tau_{ij}=2 \mu \left[ S_{ij} - \frac{1}{3} \frac{\partial u_{m}}{\partial x_{m}} \delta_{ij} \right], \qquad
q_{j}= \frac{1}{\gamma-1}\frac{\mu}{\mathrm{Pr}} \frac{\partial T}{\partial x_{j}}
\label{eq:tau_n_q}
\end{eqnarray}
where $S_{ij}$ is the strain-rate tensor, $\mu$ the dynamic viscosity, respectively given by $Sw_{ij} = \frac{1}{2}\left(\frac{\partial u_{j}}{\partial x_{i}} + \frac{\partial u_{i}}{\partial x_{j}}\right)$, $\mu = T^n$ where $n=0.75$ is the viscosity exponent and $Pr$ is the Prandtl number.

The complete set of governing equations \eqref{eq:NS_continuity}, \eqref{eq:NS_momentum}, \eqref{eq:NS_energy} are only solved for compressible turbulent channel flow with impermeable, isothermal walls in discretized form on a Cartesian domain and solved with the high-order structured code \textit{Hybrid}, originally developed by Johan Larsson for numerical investigation of the fundamental canonical shock-turbulence interaction problem \cite{LarssonL_PoF_2009,LarssonBL_JFM_2013}. The code has been scaled up to $1.97$ million cores in recent remarkable computational efforts by Bermejo-Moreno \emph{et al.} \cite{BermejoBLB_IEEE_2014}. A solution-adaptive strategy is employed to blend high-order central polynomial and WENO schemes in presence of strong flow gradient~\cite{Adams_JCP_1996,Pirozzoli_JCP_2002}.  A forth order discretization both in space and time has been adopted in the present work. Results from the high-fidelity simulations are used to generate the base flow (see section \ref{sec:base_flow}) for the present linear stability analysis.

\subsection{Linearized Equations} \label{subsec:LST_equations}

Decomposing a generic instantaneous quantity, $a(x,y,z)$ into a base state, $\mathcal{A}(y)$, and a two-dimensional fluctuation, $a'(x,y,t)$, assuming ideal gas and retaining only the first order fluctuations yields the following equations for conservation of mass, streamwise and wall-normal momentum, and energy \cite{Malik_JCP_1990}:

\begin{equation} \label{eq:linearized_mass}
\frac{\gamma}{\mathcal{T}}\frac{\partial {p'}}{\partial t}-\frac{1}{\mathcal{T}^2}\frac{\partial {T'}}{\partial t}+\frac{1}{\mathcal{T}}\frac{\partial {u'}}{\partial x}+\mathcal{U} \left( \frac{\gamma}{\mathcal{T}}\frac{\partial {p'}}{\partial x}-\frac{1}{\mathcal{T}^2}\frac{\partial {T'}}{\partial x} \right) +\frac{1}{\mathcal{T}}\frac{\partial {v'}}{\partial y}-\frac{1}{\mathcal{T}^2}\frac{\partial \mathcal{T}}{\partial y}{v'} =0 
\end{equation}
\begin{equation} \label{eq:linearized_u_momentum}
\hspace*{-1.0cm} \begin{split} \left( \frac{\partial {u'}}{\partial t}+\mathcal{U}\frac{\partial {u'}}{\partial x}+{v'} \frac{\partial \mathcal{U}}{\partial y} \right) \frac{1}{\mathcal{T}} = -\frac{\partial {p'}}{\partial x}+\frac{\mathcal{M}}{\mathrm{Re}_a} \left[ l_{2} \frac{\partial^2 {u'}}{\partial x^2} + l_{1} \left( \frac{\partial^2 {v'}}{\partial x \partial y} \right) + \frac{\partial^2 {u'}}{\partial y^2} + \right. \\ \left. \frac{1}{\mathcal{M}} \frac{d \mathcal{M}}{d \mathcal{T}}\frac{d\mathcal{T}}{dy}\left( \frac{\partial {u'}}{\partial y} +\frac{\partial {v'}}{\partial x} \right) + \frac{1}{\mathcal{M}} \frac{\partial \mathcal{M}}{\partial \mathcal{T}} \left( \frac{\partial^{2} \mathcal{U}}{\partial y^2} {T'} + \frac{\partial \mathcal{U}}{\partial y} \frac{\partial {T'}}{\partial y} \right) + \frac{1}{\mathcal{M}} \frac{\partial^2 \mathcal{M}}{\partial \mathcal{T}^2} \frac{\partial \mathcal{T}}{\partial y} \frac{\partial \mathcal{U}}{\partial y} {T'} \right] 
\end{split}
\end{equation}
\begin{equation} \label{eq:linearized_v_momentum}
\begin{split} \left( \frac{\partial {v'}}{\partial t}+\mathcal{U}\frac{\partial {v'}}{\partial x}\right) \frac{1}{\mathcal{T}}= -\frac{\partial {p'}}{\partial y}+\frac{\mathcal{M}}{\mathrm{Re}_a} \left[ \frac{\partial^2 {v'}}{\partial x^2} + l_{1} \left( \frac{\partial^2 {u'}}{\partial x \partial y} \right) + l_{2} \frac{\partial^2 {v'}}{\partial y^2} + \dots \right. \\
\left. + \frac{1}{\mathcal{M}} \frac{\partial \mathcal{M}}{\partial \mathcal{T}}\left( \frac{\partial {T'}}{\partial x} \frac{\partial \mathcal{U}}{\partial y} \right) + \frac{1}{\mathcal{M}} \frac{\partial \mathcal{M}}{\partial \mathcal{T}} \frac{\partial \mathcal{T}}{\partial y} \left\lbrace l_{0} \left( \frac{\partial {u'}}{\partial x} \right)+l_{2} \frac{\partial {v'}}{\partial y} \right\rbrace \right]\end{split}
\end{equation}
\begin{equation} \label{eq:linearized_energy}
\begin{split} \left( \frac{\partial {T'}}{\partial t}+\mathcal{U}\frac{\partial {T'}}{\partial x}+{v'} \frac{\partial \mathcal{T}}{\partial y} \right) \frac{1}{\mathcal{T}}= (\gamma-1)\left[ \frac{\partial {p'}}{\partial t}+\mathcal{U}\frac{\partial {p'}}{\partial x} \right] + \dots \hspace*{3.5cm} \\
+\frac{\mathcal{M}}{\mathrm{Re}_a\mathrm{Pr}} \left[ \frac{\partial^2 {T'}}{\partial x^2}+\frac{\partial^2 {T'}}{\partial y^2} +\frac{2}{\mathcal{K}}\frac{\partial \mathcal{K}}{\partial \mathcal{T}}\frac{\partial \mathcal{T}}{\partial y}\frac{\partial {T'}}{\partial y} + \left( \frac{1}{\mathcal{K}}\frac{\partial \mathcal{K}}{\partial \mathcal{T}}\frac{\partial^2 \mathcal{T}}{\partial y^2}+\frac{1}{\mathcal{K}}\frac{\partial^2 \mathcal{K}}{\partial \mathcal{T}^2} \left( \frac{\partial \mathcal{T}}{\partial y} \right)^2 \right) {T'} \right] +\\  \dots +\frac{(\gamma-1) \mathcal{M}}{\mathrm{Re}_a} \left[ + 2\frac{\partial \mathcal{U}}{\partial y} \left( \frac{\partial {u'}}{\partial y}+ \frac{\partial {v'}}{\partial x} \right) + \frac{1}{\mathcal{M}} \frac{\partial \mathcal{M}}{\partial \mathcal{T}} \left( \frac{\partial \mathcal{U}}{\partial y} \right)^2 \right]
\end{split}
\end{equation}
%
where $l_{j}=j-2/3$, $\mathcal{M}=\mathcal{T}^n$, $\mathcal{K}=C_{p}\,\mathcal{M}/Pr$, where  $\mathcal{M}$, $\mathcal{K}$, $\mathcal{T}$ and  $\mathcal{U}$ are the base dynamic viscosity, conductivity,  temperature and streamwise velocity. In the derivation of equations \eqref{eq:linearized_mass}, \eqref{eq:linearized_u_momentum}, \eqref{eq:linearized_v_momentum}, and \eqref{eq:linearized_energy} it was assumed that $\mathcal{P} = 1/\gamma$. The base density is derived from the base pressure and temperature via the equation of state with gas constant equal to $1/\gamma$, consistently with the normalization adopted. The effect of eddy viscosity on the fluctuations is neglected.

Assuming harmonic two-dimensional fluctuations 
\begin{equation} \label{eq:fluctuation_ansatz}
a'(x,y,t) = \hat{a}(y) e^{i(\alpha x - \omega\,t)} = \hat{a}(y)\,e^{i\alpha (x-c\,t)}
\end{equation}
where $\alpha$ is the wavenumber in the streamwise direction, and $\omega = \alpha \,c$ and $c=c_r + i\,c_i$ are the complex frequency and complex wave (or phase) speed. Applying a Gauss-Lobatto-Chebyshev discretization along the wall-normal direction yields the generalized eigenvalue problem
\begin{equation} \label{eq:generalized_eigenvalue_problem}
{\bf A} \, \pmb{\Psi} = \omega {\bf B} \, \pmb{\Psi} 
\end{equation}
with $\pmb{\Psi}$ = $\left\{ \mathbf{\hat{u}}; \mathbf{\hat{v}}; \mathbf{\hat{p}}; \mathbf{\hat{T}} \right\}$ where the generic column vector $\mathbf{\hat{a}}$ is the collection of the discretized complex amplitudes of the generic fluctuating quantity, $a'$.
Four conditions at each boundary are needed to close the system \eqref{eq:generalized_eigenvalue_problem}: no slip conditions for the streamwise velocity component, $\hat{u}=0$; isothermal conditions for temperature fluctuations, $\hat{T}=0$; impedance boundary conditions  $\hat{p} = \mp R\,\hat{v}$ for $y \pm 1$ (upper and lower wall, respectively); Non-homogeneous Neumann conditions for wall-normal derivative of pressure derived by applying \eqref{eq:linearized_v_momentum} at the boundary.

Including the aforementioned boundary conditions on the left hand side of \eqref{eq:generalized_eigenvalue_problem} results in a singular ${\bf B}$ matrix. Therefore, eigenvalues obtained by directly solving the general eigenvalue problem in this form yields eigenfunctions contaminated by numerical noise, despite eigenvalues still being accurate. This problem can be solved by recasting \eqref{eq:generalized_eigenvalue_problem} into
\begin{equation} \label{eq:generalized_eigenvalue_problem_esfahanian}
{\omega}^{-1} \pmb{\Psi} = {\bf A^{-1}}\,{\bf B}\,\pmb{\Psi}
\end{equation}
and solving the eigenvalue problem directly for the matrix ${\bf C}={\bf A^{-1}}{\bf B}$ \cite{Esfahanian_PhDThesis_1991}.  While this technique considerably improves the quality of the eigenfucntions, results for high wave-numbers ($\alpha > 2$) still exhibit numerical issues such as top-down anti-symmetry, which has limited the current investigation to $\alpha \leq 2$.


The base velocity and temperature are chosen as $\mathcal{U} = \overline{u}$ and $\mathcal{T} = \overline{T}$ where $\overline{(\;)}$ indicates Reynolds-averaged quantities, which are obtained from companion high-fidelity Navier-Stokes simulations (see section \ref{sec:base_flow}). 

In the following, validation against LSA studies available in literature is shown (section \ref{sec:ValidationLSA}) and new results from the present LSA analysis are discussed (section \ref{sec:porous_walls_results}).

%% file: Results.tex
\section{Comparison Against Previous LSA}
\label{sec:ValidationLSA}

\begin{figure}[!htbp]
\includegraphics[width=\linewidth]{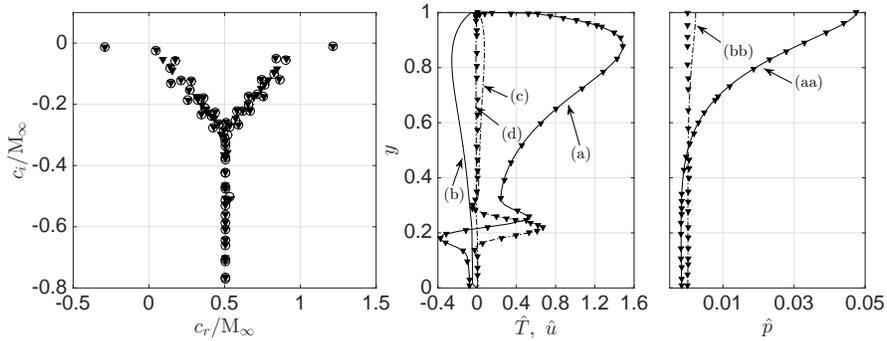}
\vspace*{-0.6cm}
\caption{Comparison of complex phase velocity spectrum for compressible Couette flow at $\mathrm{M}_{\infty}=2$, $\mathrm{Re}_{\infty}=2\times10^5$, $\alpha=1$ using $N$=100 grid points ($\circ$) with results from Hu and Zhong \cite{HuZhong_PoF_1998}  ($\blacktriangledown$). Real part (\textemdash) and imaginary part ($- \cdot -$) of velocity, $\hat{u}~$, temperature,$~\hat{T}~$, and pressure,$~\hat{p}~$, eigenfunctions of the most unstable mode at $\mathrm{M}_{\infty}=5$, $\mathrm{Re}_{\infty}=5 \times {10}^5$, $\alpha=3$ with (a): $Re \left\lbrace \hat{T} \right\rbrace $, (b): $Re \left\lbrace \hat{u} \right\rbrace$, (c): $Im \left\lbrace \hat{T} \right\rbrace$, (d): $Im\left\lbrace \hat{u} \right\rbrace$, (aa): $Re\left\lbrace \hat{p} \right\rbrace$, (bb): $Im \left\lbrace \hat{p} \right\rbrace$. }
\label{fig:compact_couette_validation_against_Hu_PoF1998}       
\end{figure}
\begin{figure}[!htbp]
\centering
\includegraphics[width=\linewidth]{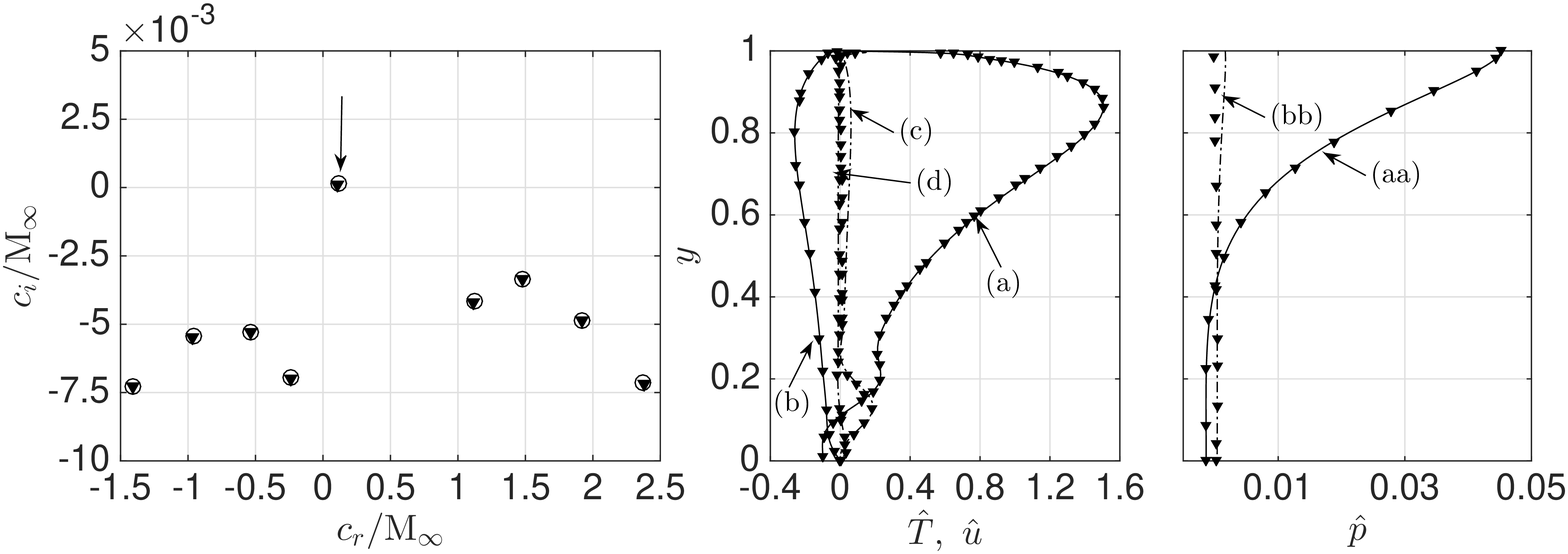}
\vspace*{-0.6cm}
\caption{Comparison of complex phase velocity spectrum for laminar compressible Couette flow at $\mathrm{M}_{\infty}=5$, $\mathrm{Re}_{\infty}=1\times{10}^{5}$ and $\alpha=2.5$ using $N$=100 grid points ($\blacktriangledown$) against Hu and Zhong \cite{Hu_AIAA_1997}. Real part (\textemdash) and imaginary part ($- \cdot -$) of velocity, $\hat{u}$, temperature, $\hat{T}$, and pressure, $\hat{p}$, eigenfunctions of the most unstable mode (indicated with downward arrow) with (a): $Re \left\lbrace \hat{T} \right\rbrace$, (b): $Re \left\lbrace \hat{u} \right\rbrace$, (c): $Im \left\lbrace \hat{T} \right\rbrace$, (d): $Im \left\lbrace \hat{u} \right\rbrace$, (aa): $Re \left\lbrace \hat{p} \right\rbrace$, (bb): $Im \left\lbrace \hat{p} \right\rbrace$. }
\label{fig:compact_couette_validation_against_Hu_AIAA1997} 
\end{figure}

Validation is first carried out against LSA results for compressible Couette flow by Hu and Zhong \cite{Hu_AIAA_1997,HuZhong_PoF_1998}, where iso-thermal conditions, $T_\infty=1$, and the tangential velocity, $U_\infty$ are imposed at the top wall, and no-slip and adiabatic conditions at the bottom wall. Reynolds number and Mach number based on the topwall velocity are $Re_\infty$ and $M_\infty$. All quantities are normalized with speed of sound based on top wall temperature, the bulk density and the total wall-to-wall distance. To enable the direct comparison with Hu and Zhong \cite{Hu_AIAA_1997,HuZhong_PoF_1998}, the dynamic viscosity have been changed to:

\begin{equation} \label{eq:HuZhongSutherlandsLaw_PrandtlNumber}
\mathcal{M} = \mathcal{T}^{1.5} \frac{1+C}{\mathcal{T}+C}, \quad C=0.5
\end{equation}
and $Pr=0.72$ (only for the sake of this comparison). Excellent agreement is found as shown by the eigenvalue spectra and eigenfunctions (figures \ref{fig:compact_couette_validation_against_Hu_PoF1998} and \ref{fig:compact_couette_validation_against_Hu_AIAA1997}). A grid convergence study of the eigenvalues (table \ref{tbl:eigenvalue_table}) also includes comparisons with more recent LSA work on the same flow by Weder \cite{Weder_Thesis_2012}.


\begin{table}
\centering
\captionof{table}{Comparison of complex wave speed, $c$, of the most unstable mode for compressible Couette flow with previously published results for various $\mathrm{Re}_{\infty}$, $M_\infty$ and $\alpha$. Grid convergence study shown for $N$=100, $N$=200 and $N$=300, respectively, top, middle and bottom rows for each case. Deviating digits from Hu and Zhong data \cite{Hu_AIAA_1997} are underlined. Values of $c$ have been normalized with the top-wall Mach number, $M_\infty$, to match published values.}
\scriptsize
\begin{tabular}{c c c c c c c c}
\label{tbl:eigenvalue_table}
\vspace*{0.05cm} \\
\multicolumn{2}{c}{Hu and Zhong \cite{Hu_AIAA_1997}} &  & \multicolumn{2}{c}{Weder \cite{Weder_Thesis_2012}} & & \multicolumn{2}{c}{Current Study}  \\ 
\cline{1-2} \cline{4-5} \cline{7-8}
 $c_{r}/M_\infty$   &   $c_{i}/M_\infty$ &  & $c_{r}/M_\infty$  &   $c_{i}/M_\infty$   & &  $c_{r}/M_\infty$   &   $c_{i}/M_\infty$ \\ 
\vspace*{0.05cm} \\
\multicolumn{8}{l}{{ $\mathrm{Re}_{\infty}=5\times{10}^{6}$, $\mathrm{M}_{\infty}=5$, $\alpha=2.1$, Mode I}} \\
\noalign{\smallskip}
+0.972869314676 & -0.003456356315 & & +0.972869\underline{178324} & -0.00345635\underline{8661} & & +0.972869\underline{280518} &-0.0034563\underline{23886} \\
+0.972869272448 & -0.003456466520 & & +0.972869\underline{198693} & -0.003456\underline{318849} & & +0.97286927244\underline{5} & -0.003456466\underline{743} \\
+0.972869272450 & -0.003456466522  & & +0.9728692\underline{01921} & -0.003456\underline{324875} & & +0.972869272\underline{355} &-0.0034564665\underline{05} \\
\vspace*{0.05cm} \\
\multicolumn{8}{l}{{ $\mathrm{Re}_{\infty}=5\times{10}^{6}$, $\mathrm{M}_{\infty}=5$, $\alpha=2.1$, Mode 0}} \\
\noalign{\smallskip}
+0.040730741952 & +0.000876050503 & & +0.0407\underline{26205831} & +0.0008\underline{84153855} & & +0.040730\underline{596292} & 	+0.00087\underline{4754264} \\
+0.040722854287 & +0.000885530891 & & +0.04072285\underline{3306} & +0.00088553\underline{1375} & & +0.040722854\underline{373} & +0.000885530\underline{566} \\
+0.040722853034 & +0.000885531421 & & +0.040722853\underline{219} & +0.000885531\underline{398} & & +0.04072285303\underline{2} & +0.000885531421 \\
\vspace*{0.05cm} \\
\multicolumn{8}{l}{{ $\mathrm{Re}_{\infty}=2\times{10}^{5}$, $\mathrm{M}_{\infty}=2$, $\alpha=0.1$, Mode I}} \\
\noalign{\smallskip}
+1.213965119859 & -0.011585118523 & & +1.21396511985\underline{1} &-0.0115851185\underline{49} & & +1.21396511985\underline{1} &-0.0115851185\underline{47} \\
+1.213965119817 & -0.011585118448 & & +1.2139651198\underline{52} &-0.011585118\underline{548} & & +1.2139651198\underline{52} & -0.011585118\underline{547} \\
+1.213965119854 & -0.011585118558 & & +1.21396511985\underline{1} &-0.0115851185\underline{48} & & +1.21396511985\underline{2} & -0.0115851185\underline{47} \\
\vspace*{0.05cm} \\
\multicolumn{8}{l}{{ $\mathrm{Re}_{\infty}=2\times{10}^{5}$, $\mathrm{M}_{\infty}=2$, $\alpha=0.1$, Mode 0}} \\
\noalign{\smallskip}
-0.291572925106  & -0.013821128462 & & -0.2915729251\underline{10} & -0.01382112846\underline{5} & & -0.29157292510\underline{9} &-0.01382112846\underline{4} \\
-0.291572925140  & -0.013821128536  & & -0.2915729251\underline{16} & -0.013821128\underline{473} & & -0.2915729251\underline{09} &-0.013821128\underline{464} \\
-0.291572925108  & -0.013821128457  & & -0.2915729251\underline{12} & -0.0138211284\underline{67} & & -0.29157292510\underline{9} &-0.0138211284\underline{64} \\
\end{tabular}
\end{table}

\begin{figure}[!h]
\includegraphics[width=\linewidth]{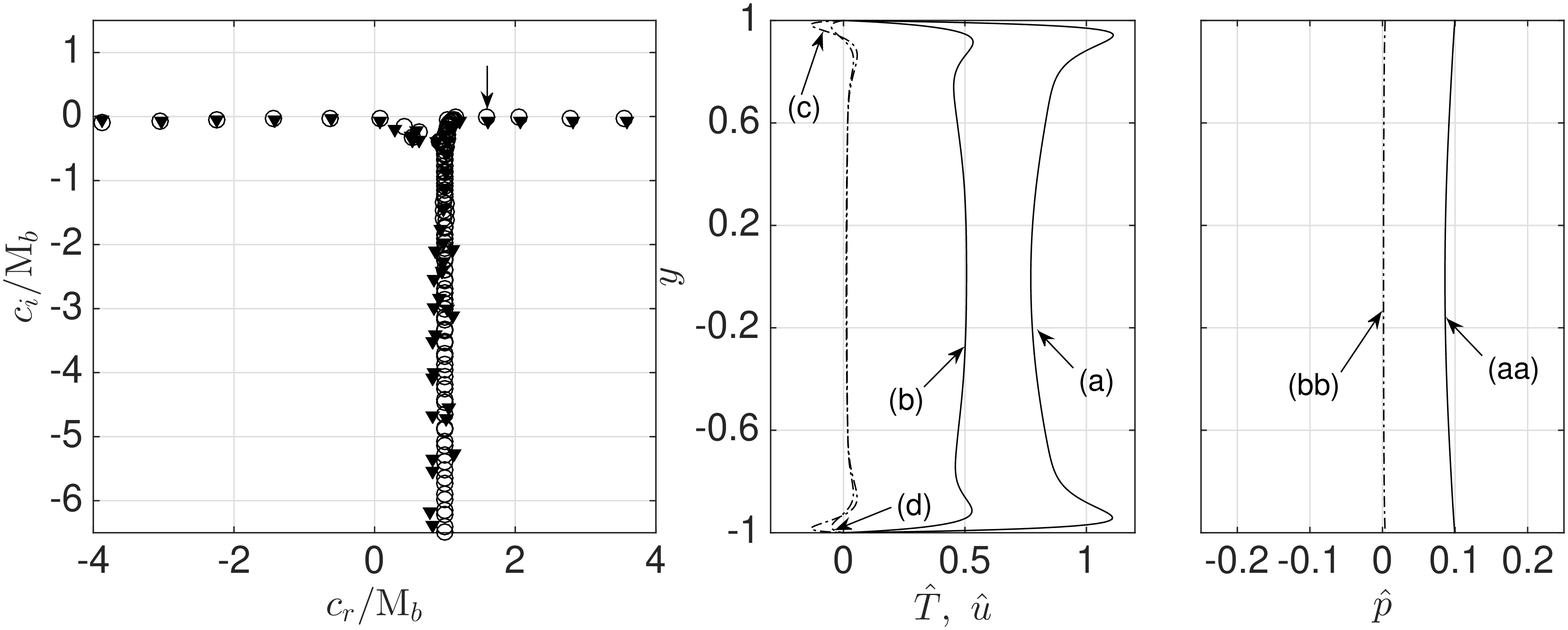}
\vspace*{-0.7cm}
\caption{Complex phase velocity spectrum for compressible channel flow at  $\mathrm{M}_{b}=0.5$, $\mathrm{Re}_{c}=4880$ and $\alpha=1$ using $N=165$ grid points ($\circ$) compared with data from Friedrich and Bertolotti \cite{Friedrich_ASR_1997} ($\blacktriangledown$). Real part (\textemdash) and imaginary part ($- \cdot -$) of eigenfunctions of the most unstable mode indicated with a downward arrow. (a): $Re \left\lbrace \hat{T} \right\rbrace$, (b): $Re\left\lbrace \hat{u} \right\rbrace$, (c): $Im\left\lbrace \hat{T} \right\rbrace$, (d): $Im\left\lbrace \hat{u} \right\rbrace$, (aa): $Re \left\lbrace \hat{p} \right\rbrace$, (bb): $Im \left\lbrace \hat{p} \right\rbrace$}
\label{fig:compact_channel_validation_against_Friedrich_ASR1997_eval}       
\end{figure} 



Validation has also been carried out against Friedrich and Bertolotti \cite{Friedrich_ASR_1997}, who performed a stability analysis of (impermeable isothermal wall) compressible turbulent channel flow at $\mathrm{Re}_{c}$=4880 based on the centerline velocity, $\mathrm{M}_{b}=3$, $\alpha=1$ using Parabolized Stability Equation (PSE) with a multi-domain spectral discretization method. The base flow was taken from the Reynolds-averaged velocity and temperature profiles in Coleman et al. \cite{ColemanKM_JFM_1995}. The eigenvalue spectrum is compared in figure \ref{fig:compact_channel_validation_against_Friedrich_ASR1997_eval} where eigenfunctions of the most unstable mode (not originally plotted in their paper) are also included. Despite the fundamental difference between our (classic) LSA approach and their PSE, the agreement is still acceptable. Friedrich and Bertolotti \cite{Friedrich_ASR_1997} did not explicitly mention what law for viscosity has been used, therefore the results in figure \ref{fig:compact_channel_validation_against_Friedrich_ASR1997_eval} have been reproduced assuming $\mu = \mathcal{T}^n$, with $n=0.7$, consistent with the simulations by Coleman et al. \cite{ColemanKM_JFM_1995}.

\section{Compressible Channel Flow with Porous Walls} \label{sec:porous_walls_results}

In the following results from impermeable-wall compressible turbulent channel flow high-fidelity simulations are first discussed (section \ref{sec:base_flow}). The mean velocity and temperature profiles extracted from the latter are used for the present linear stability analysis with porous walls (section \ref{sec:LSA_channel_porous}).

\subsection{Impermeable-Wall Turbulent Channel Flow} \label{sec:base_flow}

The base velocity, density and temperature profiles for the present LSA investigations have been taken from laminar and turbulent numerical simulations of fully compressible impermeable isothermal-wall channel flow.  

The computational domain considered for the turbulent simulations in this study is $L_{x} \times L_{y} \times L_{z} = 8 \times 2 \times 4$ which is discretized with a number of control volumes $N_{x} \times N_{y} \times N_{z}=256 \times 128 \times 192$ resulting in a quasi-DNS resolution of $\Delta x^+ \sim 12.27-13.22$, $\Delta z^+ \sim 8.17-8.81$, ${\Delta y^+}_\textrm{min} = 0.45-0.47$, over the range of bulk Mach numbers investigated. The superscript $^+$ indicates classic wall-units, $\delta_\nu = u_\tau \rho_w/\mu_w$, where $u_\tau=\sqrt{\tau_w/\rho_w}$ is the friction velocity and $\rho_w$ and $\mu_w$ are the density and dynamic viscosity evaluated at the wall and the wall-shear stress is ${\tau}_{w}={\mu}_{w}~{\partial \overline{u}}/{\partial y}\vert_{w}$ where $\overline{(\;)}$ indicates Reynolds averaging.


\begin{figure}[!htbp]
\includegraphics[width=\linewidth]{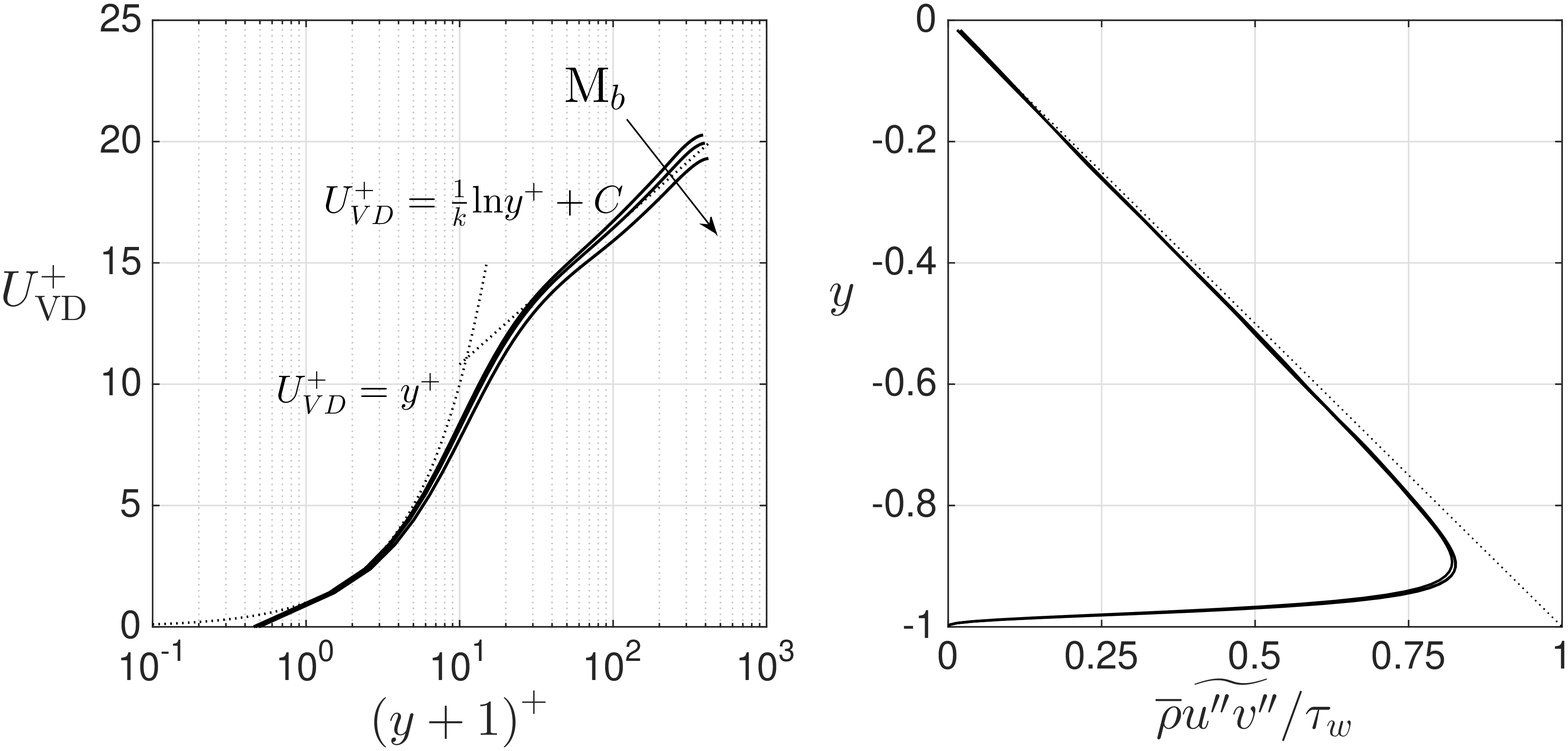}
\put(-330,5){(a)}
\put(-10,5){(b)}
\caption{Profiles of turbulent mean velocity following Van Driest transform with reference law of the wall and log-law $k=0.41$ and $C=5.2$ ($\cdot\cdot\cdot$) (a) and Reynolds stresses normalized by the average wall shear stress, $\tau_{w}$ (b)}
\label{fig:semilocal_channel_6900}
\end{figure}

Mean velocity profiles in figure (\ref{fig:semilocal_channel_6900} a) are plotted using Van Driest transform following
\begin{equation}\label{eq:VD}
{U}^{+}_{VD}=\int_{0}^{U^+} { \left( \frac{\rho}{\rho_{w}} \right)^{1/2} d U^{+}}=\frac{1}{\kappa} \ln{y^+}+C
\end{equation}
In this set of simulations, friction Reynolds number, $\mathrm{Re}_{\tau}=u_{\tau}\,\delta/\nu$, varies from 385.6 to 417.4. Collapse of both mean velocity and Reynolds Stress profiles is obtained, showing that the state of near-wall turbulence has been successfully kept constant while varying the bulk Mach number.

\begin{figure}
\includegraphics[width=\linewidth]{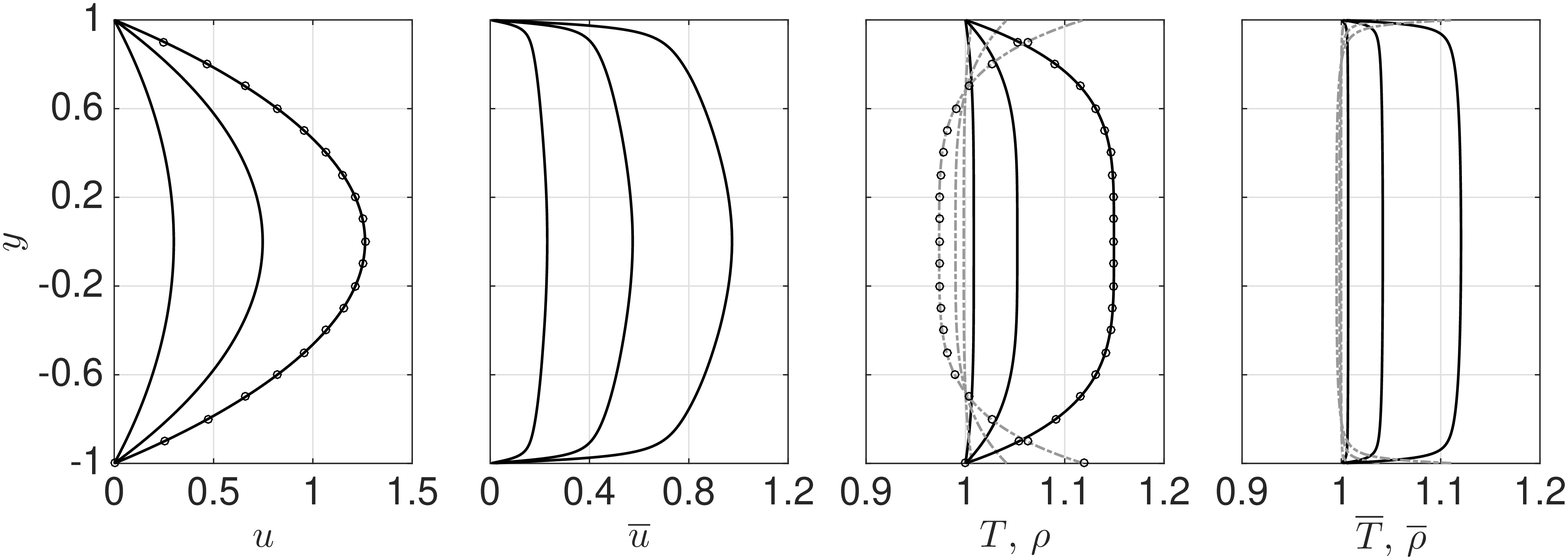}
\caption{Profiles of laminar velocity, $u(y)$, temperature, $T(y)$, and density $\rho(y)$ and Reynolds-averaged turbulent velocity, $\overline{u}(y)$, temperature, $\overline{T}(y)$, and density $\overline{\rho}(y)$ at $\mathrm{Re}_{b}=6900$ and $M_{b}$ = 0.2, 0.5, and 0.85. Symbols show laminar results calculated at $\mathrm{Re}_{b}=6900$ and $M_{b}$ = 0.85 obtained separately by discretizing the governing equations retaining only the streamwise velocity component and wall-normal gradients.}
\label{fig:mean_profile_channel_ttight_re=6900}
\end{figure}

The laminar base flow has been generated with the same code with $N_{y}=192$ control volumes in the wall-normal direction, for values of $Re_b$ and $M_b$ matching the turbulent simulations, and are shown in figure \ref{fig:mean_profile_channel_ttight_re=6900}.

\subsection{Linear Stability Analysis With Porous Walls} \label{sec:LSA_channel_porous}

In the following we explore the effects of porous walls on the linear stability of the two-dimensional perturbation \eqref{eq:fluctuation_ansatz} using the impermeable channel flow results generated in section \ref{sec:base_flow} as the base flow. Since turbulent eddy viscosity is neglected in \eqref{eq:linearized_u_momentum}, \eqref{eq:linearized_v_momentum}, and \eqref{eq:linearized_energy}, the nature of the instability between laminar and turbulent cases only differs due to the base flow. In particular, the profile of the wall-normal gradient of the mean streamwise velocity plays a fundamental role in shaping the unstable modes triggered by the wall permeability and determining the location of the energy production regions. In the following, the trajectory of the eigenvalues of the dominant unstable modes is analyzed, together with their induced excess Reynolds shear stresses.

\subsubsection{Unstable Modes Eigenvalues Trajectory}\label{subsec:Eval_trajectory}

\begin{figure}[!ht]
\centering
\includegraphics[width=0.95\linewidth]{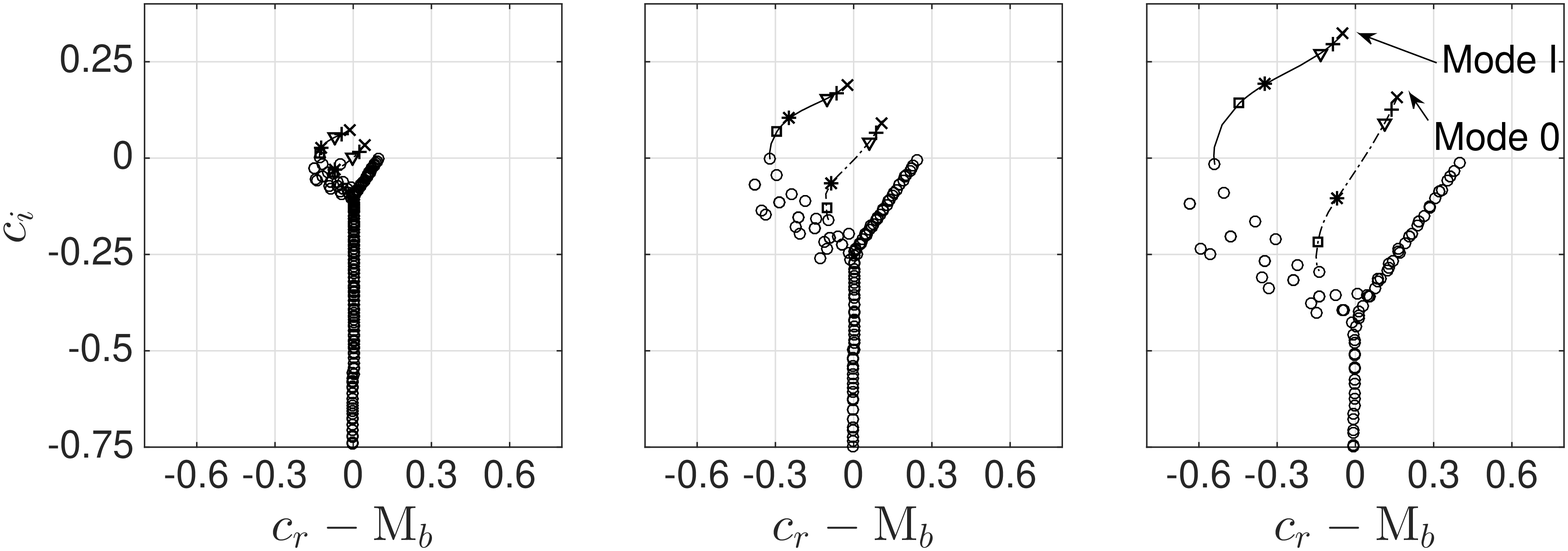}
\put(-320,10){(a)}
\put(-255,115){$M_b=0.2$} 
\put(-160,115){$M_b=0.5$}
\put(-65,115){$M_b=0.85$} \\
\includegraphics[width=0.95\linewidth]{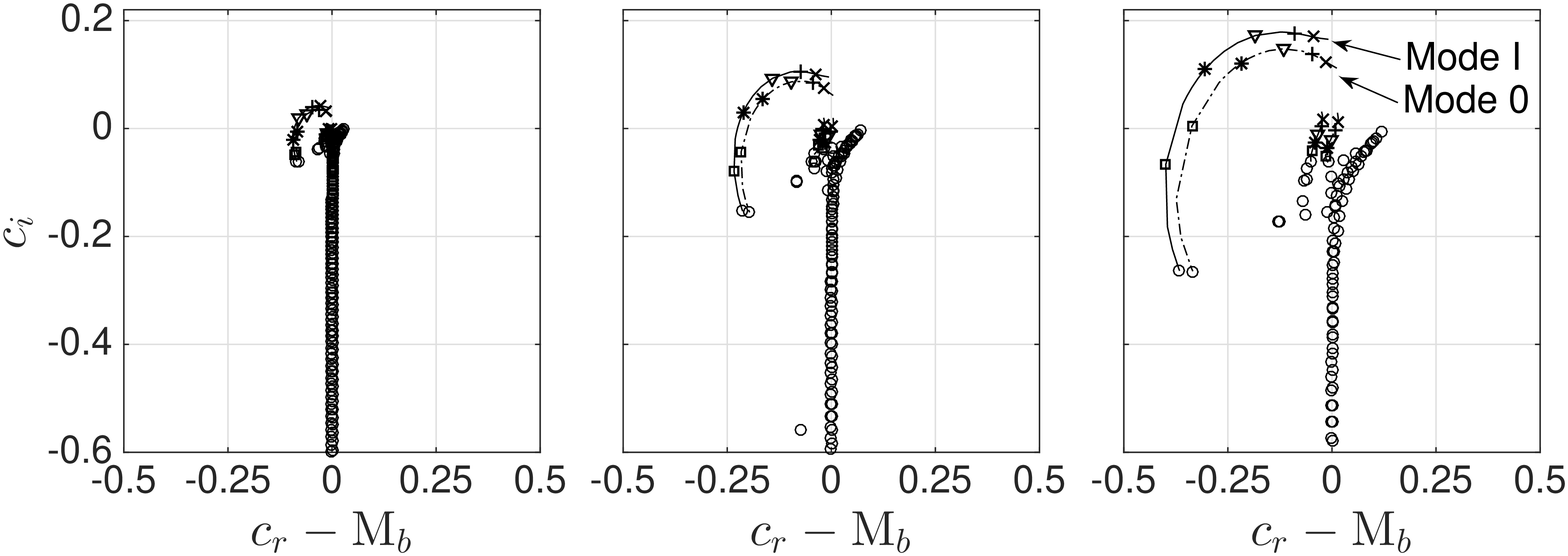}
\put(-320,10){(b)}
\caption{Trajectory of Mode 0 and Mode I for $\alpha$=1 in the complex phase velocity spectrum of compressible channel flow with impedance boundary conditions \eqref{eq:purely_real_IBC} using laminar (a) and turbulent (b) base flow at $Re_{b}=6900$, $\mathrm{M}_{b}=0.85$ traced by varying the impedance resistance in the range $R=10 - 0.01$ for $N=200$ grid points. In increasing order of permeability, $R\to\infty$ (zero permeability): ($\circ$), $R=1: \Box$, $R=0.5: \ast$, $R=0.1: \triangledown$, $R=0.05: +$, $R=0.01: \times $.}
\label{fig:compact_channel_trajectory}       
\end{figure}

A parametric study has been performed by varying the impedance resistance, $R$, from $R \to \infty$ (zero permeability) to $R=0.01$ (high permeability). Two modes, Mode 0 and Mode I, are made unstable ($c_i>0$) for a sufficiently low value of resistance $R<R_\textrm{cr}$ (figure \ref{fig:compact_channel_trajectory}) for both laminar and turbulent base flows. In all cases, the phase speed of the unstable modes increases monotonically with increasing the permeability, exceeding the bulk velocity. 

In the laminar case, for $\alpha=1$, the two modes originate from different regions of the eigenvalue spectrum. Mode I at $R\to\infty$ exhibits a very slow decay rate, requiring low values of permeability (high value of resistance $R_{cr}=1$) to become unstable. For near-impermeable conditions, Mode 0 is instead initially located close to the junction of Y-shape spectrum (phase velocity approximately equal to the bulk velocity) and becomes unstable at much lower values of resistance $R_\textrm{cr} \sim 0.05$ (higher values of permeability). However, the differences in their instability dynamics and spatial structures (discussed later) are not due to their different location in the eigenvalue spectrum.
%
As $\alpha$ increases (not shown), the two starting points, and the corresponding trajectories, become closer such that for $\alpha=2$ they start from two adjacent nodes at the middle of the left branch of Y-spectrum. Decreasing $\alpha$ yields the opposite effect, with trajectories originating from opposite ends of the Y spectrum for $\alpha=0.1$.

The impermeable-wall spectrum for $\alpha=1$ does not exhibit a clear Y-shape for the turbulent case. The trajectories of the modes follow very similar paths in the range of $\alpha=0.1 - 2$ (not shown) tested and share very similar values of critical impedance resistance (or wall permeability), suggesting their coexistence. This behavior is due to the \emph{flatness} of the mean velocity profile, which is even more accentuated in the case of fully developed turbulence over tuned wall impedance, as observed by Scalo et al. \cite{ScaloBL_PoF_2015}.


The growth rates of the unstable modes, in case of laminar base flow, increase monotonically for decreasing $R$ in the range investigated. However, when using turbulent base flow, the growth rate saturates and then decreases for the same range of $R$. For the turbulent case only, two new unstable modes (with a similar pairing behavior as Mode 0 and Mode I) originate from the central part of the spectrum. They are not present in the laminar base flow case, travel at roughly the bulk velocity, ($c_r \sim M_b$) and become unstable, for very low values of resistance $R_{cr} \sim 0.05$. A detailed study of them is deferred to future work.

\begin{figure}[!htbp]
\centering
\includegraphics[width=0.9\linewidth]{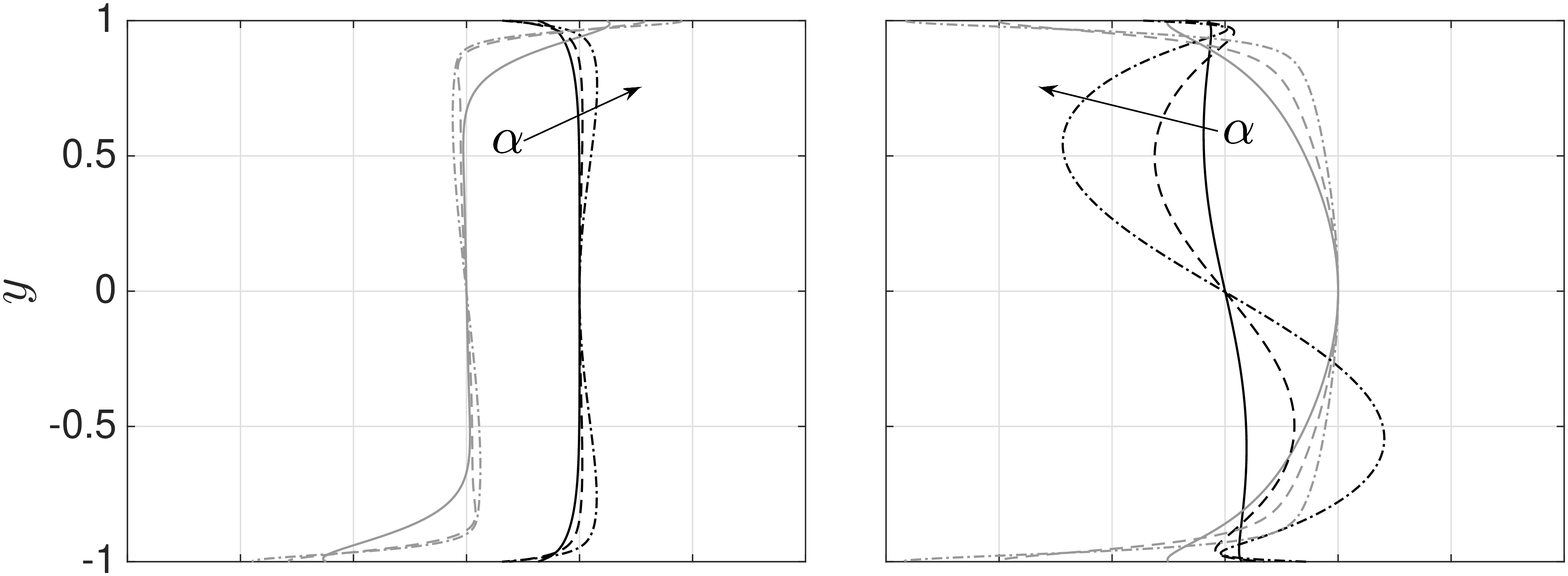}
\put(-215,112){Mode 0}
\put(-75,112){Mode I}
\put(-300,5){(a)}
\vspace{0.2cm}
\includegraphics[width=0.9\linewidth]{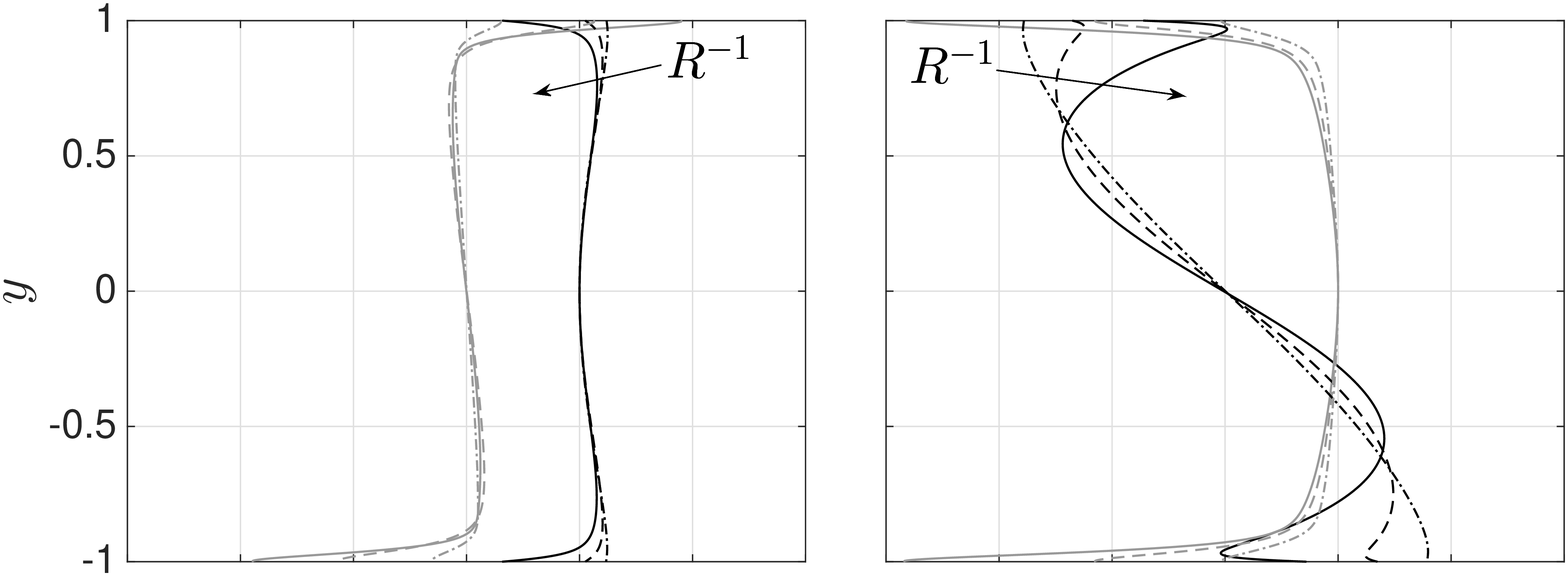}
\put(-300,5){(b)}
\vspace{0.2cm}
\includegraphics[width=0.9\linewidth]{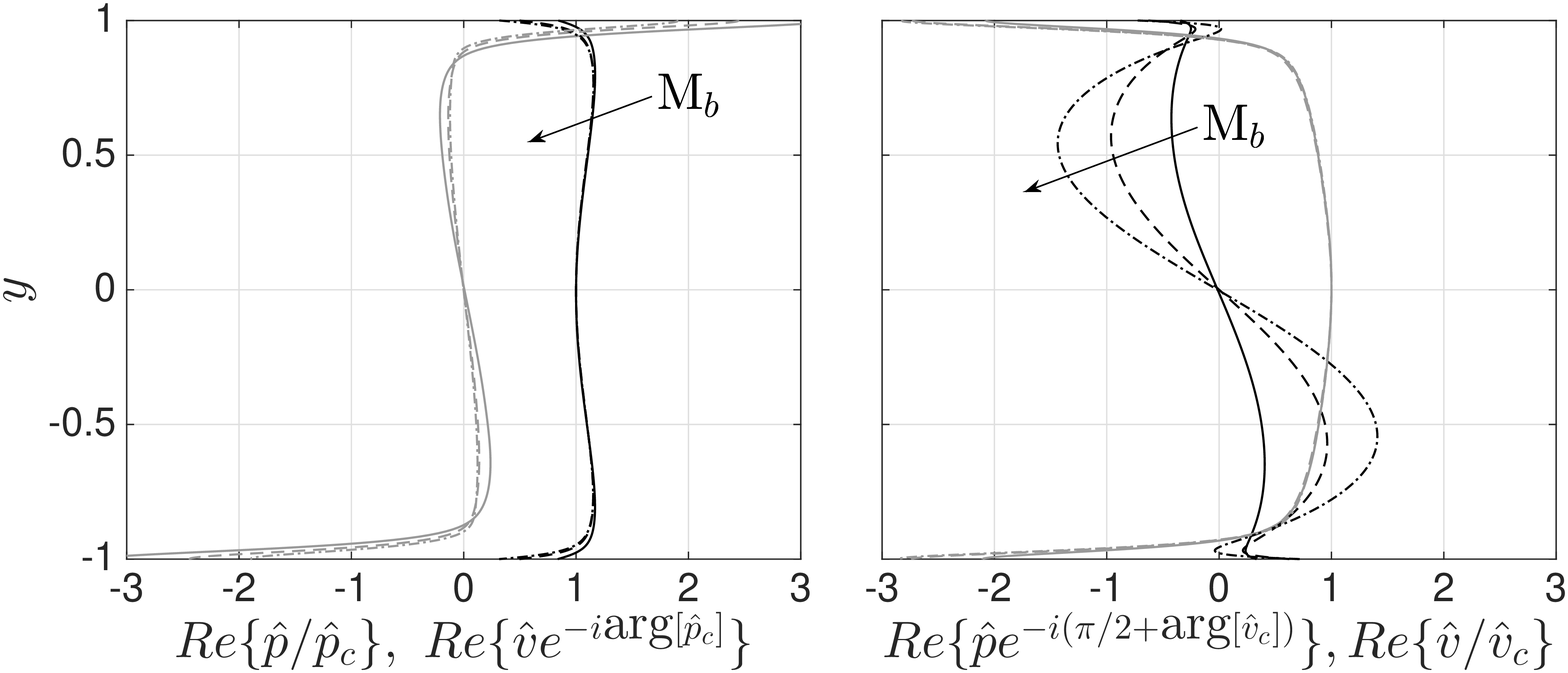}
\put(-300,0){(c)}
\caption{Comparison between Mode 0 (left) and Mode I (right) showing pressure, $\hat{p}$, (black) and wall-normal velocity, $\hat{v}$ (grey) eigenfunctions with base flow taken from turbulent mean profiles at $\mathrm{Re}_b = 6900$. Resistance $R=0.01$ and $\mathrm{M}_b$ = 0.85, for wavenumbers $\alpha = 0.1$ (---), $\alpha=0.5$ (- - -), and $\alpha=1.0$ ($-\cdot-$) (a); Wavenumber $\alpha=1$ and $\mathrm{M}_b$ = 0.85, for resistances $R =1.0$ ($-\cdot-$), $R=0.1$ (- - -), and $R = 0.01$ (---) (b). Wavenumber $\alpha=1$ and $R$ = 0.01, for bulk Mach numbers $M_b = 0.2$ (---), $M_b = 0.5$ (- - -), and $M_b = 0.85$ ($-\cdot-$) (c). In all Mode 0 plots, all pressure eigenfunctions have been normalized with their value at the centerline, $\hat{p}_c$; the resulting phase shift has been applied to the eigenfunctions of wall-normal velocity. In all Mode I plots, the vertical velocity eigenfunctions have been normalized with their value at the centerline, $\hat{v}_c$, the pressure eigenfuctions' phase has been shifted accordingly and then further changed by 90$^{\circ{}}$ to highlight the standing-wave-like structure in the channel core.}
\label{fig:eigenfunction_compare_k_R_Mb} 
\end{figure}

The transition from subcritical to supercritical permeability does not qualitatively alter the structure of Mode 0 and Mode I; both modes retain a very pronounced acoustic-like structure in the core of the channel (figure \ref{fig:eigenfunction_compare_k_R_Mb}), even in the impermeable wall limit.
For the range of $\alpha$ and Mach numbers investigated, Mode 0 always manifests as a bulk pressure mode, causing symmetric expulsion and suction of mass from the porous walls, whereas Mode I resembles a standing wave, with a pressure node at the channel centerline. The only exception is Mode 0 in the case of laminar base flow profiles at subcritical permeabilities, which does not exhibit an appreciable structural coherence.

Increasing the wavenumber $\alpha$ results in an increase of relative intensity of the wall-normal transpiration with respect to the centerline value of $|\hat{v}|$ for both modes (figure \ref{fig:eigenfunction_compare_k_R_Mb}a) and a thinning of the boundary layer of the perturbation, extending the inviscid acoustic-like core region. Future studies will verify if such effects are consistent with a reduction of the Stokes boundary layer thickness, $\delta_s = \sqrt{2\nu/\omega_r} = \sqrt{2\nu/(\alpha\,c_r)}$.

As permeability increases, the shape of Mode I deviates from a typical standing-wave-like structure, with two new pressure nodes appearing in the near wall region. An increase in the relative intensity of the wall-normal transpiration with respect to $\hat{v}_{c}$ in Mode 0 (figure \ref{fig:eigenfunction_compare_k_R_Mb}b) is also observed. Varying bulk Mach number, in the range considered, does not significantly alter the structure of both modes (figure \ref{fig:eigenfunction_compare_k_R_Mb}c).

\subsubsection{Perturbation-Induced Reynolds Shear Stress}

The modifications to the Reynolds Shear Stress (RSS) distribution that would result from the application of porous walls can be qualitatively predicted by analyzing the normalized, perturbation-induced excess shear stresses
\begin{equation} \label{eq:reynolds_stress_shape}
\widetilde{R}_{12} = \frac{Re\left\{\hat{u} \hat{v}^*\right\}}{\int_{-1}^0 Re\left\{\hat{u} \hat{v}^*\right\} dy}.
\end{equation}
In the case of turbulent mean flow profiles, both modes generate an anti-symmetric RSS distribution concentrated in base flow's viscous sublayer (figure \ref{fig:semilocal_channel_6900}b) with negligible effects in the turbulent core ($-0.9<y<0.9$).

\begin{figure}[!ht]
\centering
\includegraphics[width=0.9\linewidth]{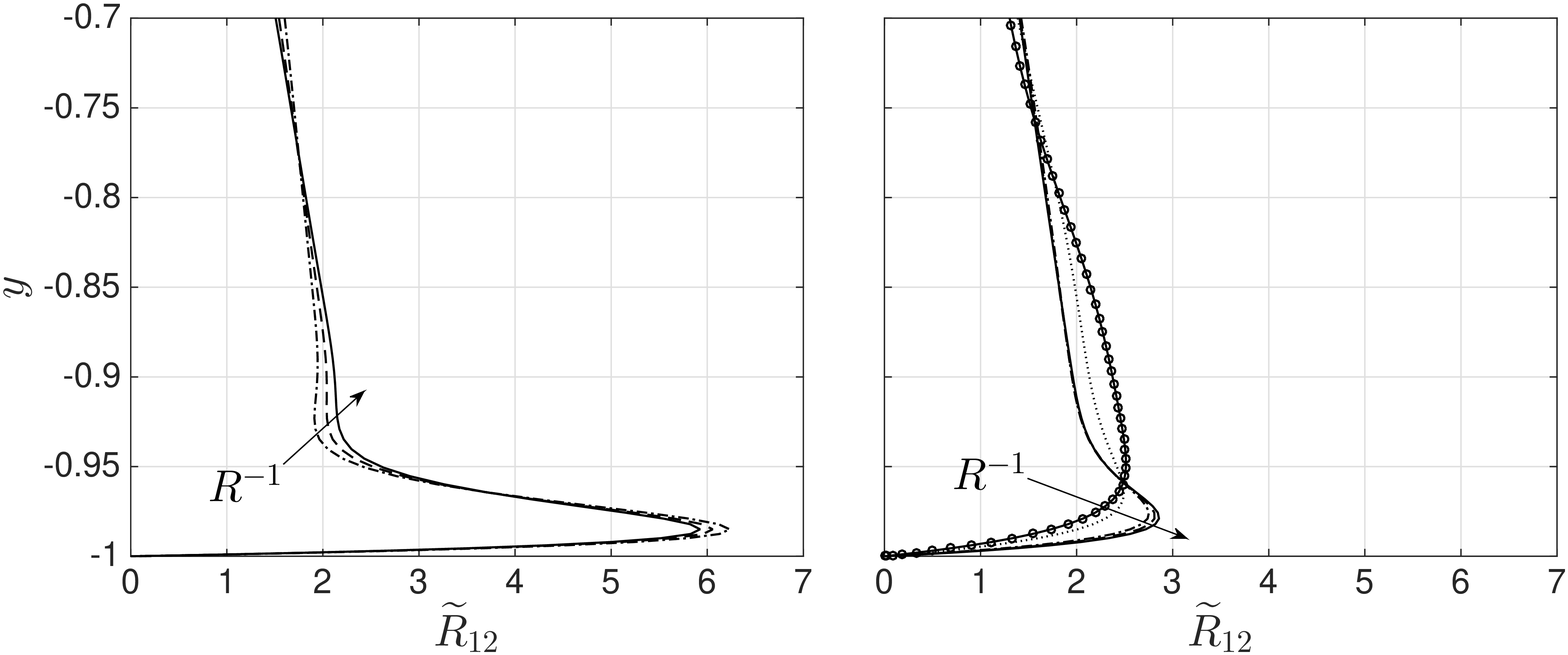}
\put(-300,10){(a)}
\put(-230,130){ Mode 0}
\put(-80,130){Mode I}
\\
\includegraphics[width=0.9\linewidth]
{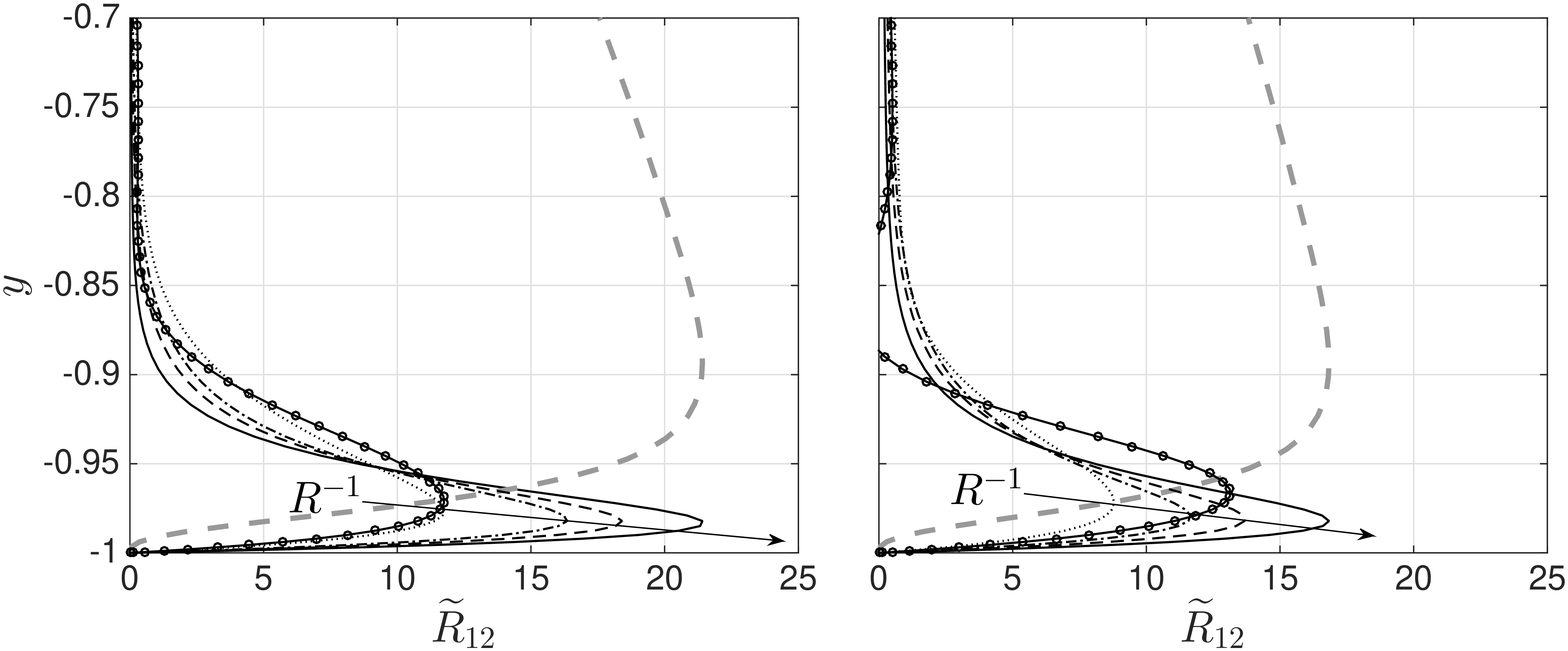}
\put(-300,10){(b)}
\vspace*{-0.1cm}
\caption{Near-wall profiles of normalized perturbation-induced RSS, $\widetilde{R}_{12}$ \eqref{eq:reynolds_stress_shape} for Mode 0 (left) and Mode I (right) predicted by LSA for $N=$200 grid points, for different values of resistance, $R$, and $\alpha=1$ with base flow taken from laminar (a) and turbulent (b) channel flow calculations at $\mathrm{Re}_{b}=6900$, $\mathrm{M}_{b}=0.85$. Arrows indicate increasing values of permeability (decreasing value of resistance) with $R = 1$ ($\circ$),  $R=0.5$ ($\cdot \cdot$), $R= 0.1$ ($- \cdot - \cdot$), $R= 0.05$ (- - -), $R=0.01$ (\textemdash). Results for R=1 and R=0.5 are omitted for Mode 0 with a laminar base profile, since corresponding eigenmodes do not exhibit a coherent structure. Turbulent RSS from the impermeable wall calculations (figure \ref{fig:semilocal_channel_6900}) are reported here (in arbitrary units) with thick dashed grey lines.}
\label{fig:compact_channel_reynoldsstress}       
\end{figure}

When employing a laminar base flow, normalized RSS for Mode 0 using laminar base flow do not change considerably as permeability varies. However, the corresponding eigenmodes are structurally incoherent for subcritical permeabilities, as also pointed out in section \ref{subsec:Eval_trajectory}. Results for Mode I show that its RSS peak approaches the wall for increasing the permeability (decreasing $R$ value). With a turbulent base flow, the RSS shape for both modes exhibits very similar features. Decreasing the $R$ value brings the RSS peak closer to the wall with normalized values evanescent in the channel core (in contrast with the laminar base flow results). It should be noted that increasing the permeability (lowering value of $R$) results in a higher growth rate of the disturbance, and will therefore yield a more intense augmentation of the total turbulent RSS (figure \ref{fig:compact_channel_reynoldsstress}).

The predicted shape of the perturbation-induced RSS profiles is consistent with the structure of the Kelvin-Helmholtz instability observed by Scalo et al \cite{ScaloBL_PoF_2015}, sustained by the wall-normal acoustic resonance generated by the assigned wall impedance.

%% file: Conclusions.tex
\section{Discussion and Future Work}

In the present paper, Linear Stability Analysis (LSA) is employed to predict the effects of wall-permeability on a laminar and turbulent compressible channel flow up to transonic bulk Mach numbers, for a fixed bulk Reynolds number of $Re_{b}=6900$. Two-dimensional streamwise-periodic disturbances are assumed and porous walls are modeled by linear acoustic impedance boundary conditions (IBC) with zero reactance. This results in a purely real impedance, defined uniquely by its resistance, $R$, inversely proportional to the wall-permeability. The resistance has been varied in the range, $R=1 -0.01$.

Increasing the permeability, or equivalently the inverse of the resistance $R^{-1}$, destabilizes the flow by triggering two modes; one characterized by bulk (quasi-uniform) pressure oscillations in the core, with gradients of fluctuating quantities concentrated in the near wall viscous region (Mode 0); the other resembling an acoustic standing wave resonating in the channel's core, with similar near-wall behavior (Mode I). Mode 0 is characterized by symmetric $u'$, $T'$, and $p'-$ eigenmodes and anti-symmetric $v'$. The latter is associated with an anti-symmetric, periodic, expulsion and suction of mass from the boundaries driven by the bulk pressure fluctuations in the core. On the other hand, Mode I exhibits anti-symmetric $u'$, $T'$, and $p'$ profiles with symmetric $v'$ distribution. For both modes, increasing the wall permeability causes the perturbation-induced Reynolds shear-stress peak to shift closer to the porous walls, for both laminar and turbulent base flows.

For very high values of permeability (low values of resistance $R<0.01$) and only in the case of a turbulent base flow, a saturation of the growth rate is observed. Values of $R<0.1$ 
may, however, be unrealistic (see measurements of the acoustic response of perforated panels \cite{TamA_AIAA_1996}) and are also expected to yield nonlinear effects that are not captured by a linear IBC formulation.

Although the current analysis is limited to purely real acoustic impedances (i.e. classic Darcy-like formulation for porous walls), the results obtained qualitatively confirm the structure of the hydro-acoustic instability observed by Scalo \emph{et al.} \cite{ScaloBL_PoF_2015}. In their case, large spanwise-coherent Kelvin-Helmholtz rollers are found to be confined near the wall where Reynolds shear stresses are significantly augmented by the instability, while the outer layer remained unaffected. This result is in qualitative agreement with the perturbation-induced Reynolds shear stress distribution predicted by the current analysis and in contrast with the findings by Jim\'enez et al. \cite{Jimenez2001Turbulent}, valid for the incompressible limit, where Kelvin-Helmholtz rollers were observed in the outer layer and the near wall turbulence was unaltered. A more accurate comparison between our LSA and the results by Scalo \emph{et al.} \cite{ScaloBL_PoF_2015} will be obtained in future studies, which will incorporate a broadband reactance (the impedance's imaginary part) in the eigenvalue problem.

Eigenvalue tracking shows that the modes made unstable by permeability preserve the characteristic structure they had when stable. Only in the laminar case, Mode I is at the verge of instability for zero permeability. As the wavenumber $\alpha$ is increased, trajectory of the unstable modes become close to each other, and near-wall gradients of fluctuating quantities intensify.

Future work will focus on Direct Numerical Simulation of channel flow with general impedance boundary conditions and new companion LSA efforts aimed at finding the optimal impedance distribution for effecting flow control, with improved numerics and physical modeling of the boundary conditions. Investigations will be also carried out considering boundary layers over flat plates, to rule out confinement effects.

\section{From Small Data to Big Data (and Vice Versa)}

This work has been centered around the use and generation of \emph{small} data, in the form of a classic Linear Stability Analysis, in order to acquire a \emph{general} understanding of the response that it is expected from compressible channel flow turbulence under the effects of porous walls. This problem can only be truly explored \emph{in detail} with large-scale, high-fidelity simulations, that is, \emph{big} data. Simplified low-order models, however, prove to be essential in guiding the design and interpretation of companion large-scale numerical simulations, since they provide us with a rough but robust understanding of the physical problem under investigation at a considerably reduced cost. Moreover, low-order models allow for a true parametric study, spanning, for example, several orders of magnitude of variation of a key parameter in the problem, such as, in this case, the wall permeability. High-fidelity simulations, on the other hand, can be used to populate a few key regions of the parameter space with limited modeling assumptions, and used to calibrate and/or verify the results from the low-order modeling efforts.

When possible, a synergistic interaction between small data and big data should therefore be sought. A ``blind big data''- or a ``blind small data''- only approach may suffer from severe limitations. The development, analysis and adoption of the two approaches should go hand in hand in order to cross-verify results, especially when experimental data is not available. However, with the advances in computational power, the definition of small and big data is changing rapidly. For example, turbulence simulations performed on a computational grid of $N^3 = 1024^3$ points, can nowadays (2015) be considered \emph{average}-size data. In spite of the gradual shift in the demarcation line between small and big data, we believe that the importance of low-order models (used here as an example of small data) will never (and should never) cease to retain a prominent role in the design, analysis and support of results from large-scale numerical investigations.

%% file: RahbariScalo_WhitherTurbulence_2015.bbl
\begin{thebibliography}{10}%
%
%
%
%
%
%
%




\bibitem{Adams_JCP_1996}
N.~A. Adams and K.~Shariff.
\newblock Conservative hybrid compact-weno schemes for shock-turbulence
  interaction.
\newblock {\em J Comput Phys}, 1996.

\bibitem{BermejoBLB_IEEE_2014}
Ivan Bermejo-Moreno, J.~Bodart, J.~Larsson, and B.M. Barney.
\newblock {Solving the compressible Navier-Stokes equations on up to 1.97
  million cores and 4.1 trillion grid points}.
\newblock In {\em IEEE International Conference on High Performance Computing},
  2013.

\bibitem{BresICF_JFM_2013}
G.~A. Bres, M.~Inkman, T.~Colonius, and A.~V. Fedorov.
\newblock {Second-mode attenuation and cancellation by porous coatings in a
  high-speed boundary layer}.
\newblock {\em J. Fluid Mech.}, 726:312, 2013.

\bibitem{ColemanKM_JFM_1995}
G.~N. Coleman, J.~Kim, and R.~D. Moser.
\newblock {A numerical study of turbulent supersonic isothermal-wall channel
  flow}.
\newblock {\em J. Fluid Mech.}, 305(-1):159--183, December 1995.

\bibitem{TullioS_PoF_2010}
N.~De~Tullio and N.~D. Sandham.
\newblock {Direct numerical simulation of breakdown to turbulence in a Mach 6
  boundary layer over a porous surface}.
\newblock {\em Phys. Fluids}, 22:094105, 2010.

\bibitem{Duck_JFM_1994}
Hussaini M~Y Duck P~W, Erlebacher~G.
\newblock On the linear stability of compressible plane couette flow.
\newblock {\em J. Fluid Mech.}, 258:131--165, 1994.

\bibitem{Esfahanian_PhDThesis_1991}
Vahid Esfahanian.
\newblock {\em Computation and stability analysis of laminar flow over a blunt
  cone in hypersonic flow}.
\newblock PhD thesis, The Ohio State University, 1991.

\bibitem{Friedrich_ASR_1997}
R.~Friedrich and F.~P. Bertolotti.
\newblock Compressibility effects due to turbulent fluctuations.
\newblock {\em Appl. Sci Res}, 57(165--194), 1997.

\bibitem{Lekoudis_AIAA_1978}
Lekoudis~S. G.
\newblock Stability of boundary layer over permeable surfaces.
\newblock In {\em AIAA Aerosp. Sci. Meet. Exhibit}, Huntsville, AL 1-8, 1978.

\bibitem{Hu_AIAA_1997}
S.~Hu and X.~Zhong.
\newblock Linear instability of compressible plane couette flows.
\newblock In {\em AIAA Aerosp. Sci. Meet. Exhibit}, pages 1--16, Reno, NV,
  1997.

\bibitem{HuZhong_PoF_1998}
Sean Hu and Xiaolin Zhong.
\newblock Linear stability of viscous supersonic plane couette flow.
\newblock {\em Phys. Fluids}, 10(3):709 -- 730, 1998.

\bibitem{jimenez1999autonomous}
J.~Jim{\'e}nez and A.~Pinelli.
\newblock {The autonomous cycle of near-wall turbulence}.
\newblock {\em J. Fluid Mech.}, 389:335--359, 1999.

\bibitem{Jimenez2001Turbulent}
J.~Jim{\'e}nez, M.~Uhlmann, A.~Pinelli, and G.~Kawahara.
\newblock {Turbulent shear flow over active and passive porous surfaces}.
\newblock {\em J. Fluid Mech.}, 442:89--117, September 2001.

\bibitem{LarssonL_PoF_2009}
J.~Larsson and S.K. Lele.
\newblock Direct numerical simulation of canonical shock/turbulence
  interaction.
\newblock {\em Phys.\ Fluids Fluids}, 21, 2009.

\bibitem{LarssonBL_JFM_2013}
Johan Larsson, Ivan Bermejo-Moreno, and Sanjiva~K. Lele.
\newblock Reynolds- and mach-number effects in canonical shock--turbulence
  interaction.
\newblock {\em Journal of Fluid Mechanics}, 717:293--321, 2 2013.

\bibitem{Mack_TCFD_1990}
Mack~L M.
\newblock On the inviscid acoustic-mode instability of supersonic shear flows.
\newblock {\em Theor. Comput. Fluid Dyn.}, 2(2):97--123, 1990.

\bibitem{Malik_PR_2012}
M.~Malik, J.~Dey, and M.~Alam.
\newblock Linear stability, transient energy growth, and the role of viscosity
  stratification in compressible plane couette flow.
\newblock {\em Phys Rev E}, 77:1--15, 2012.

\bibitem{Malik_JCP_1990}
M.~L. Malik.
\newblock Numerical methods for hypersonic boundary layer stability.
\newblock In {\em J Comput Phys}, volume~86, pages 376--413, 1990.

\bibitem{Pirozzoli_JCP_2002}
S.~Pirozzoli.
\newblock Conservative hybrid compact-weno schemes for shock-turbulence
  interaction.
\newblock {\em J. Comput. Phys.}, 178:81--117, 2002.

\bibitem{ScaloBL_PoF_2015}
C.~Scalo, J.~Bodart, and S.~K. Lele.
\newblock Compressible turbulent channel flow with impedance boundary
  conditions.
\newblock {\em Phys. Fluids}, 27(035107), 2015.

\bibitem{TamA_AIAA_1996}
C.~K.~W. Tam and L.~Auriault.
\newblock {Time-domain Impedance Boundary Conditions for Computational
  Aeroacoustics}.
\newblock {\em AIAA J.}, 34(5):917 -- 923, 1996.

\bibitem{Tilton_JFM_2008}
N.~Tilton and L.~Cortelezzi.
\newblock Linear stability analysis of pressure-driven flows in channels with
  porous walls.
\newblock {\em J. Fluid Mech.}, 604:411--445, 2008.

\bibitem{Wagner:2014PhD}
A.~Wagner.
\newblock {\em Passive Hypersonic Boundary Layer Transition Control Using
  Ultrasonically Absorptive Carbon-Carbon Ceramic with Random Microstructure}.
\newblock PhD thesis, Katholieke Universiteit, Leuven, 2014.

\bibitem{WagnerHK_AIAA_2012}
A.~Wagner, K~Hannemann, and M.~Kuhn.
\newblock Experimental investigation of hypersonic boundary-layer stabilization
  on a cone by means of ultrasonically absorptive carbon-carbon material.
\newblock AIAA Paper 2012-5865, 2012.

\bibitem{WartemannLS_AIAA_2012}
V.~Wartemann, H.~L\"{u}deke, and N.~D. Sandham.
\newblock {Numerical Investigation of Hypersonic Boundary-Layer Stabilization
  by Porous Surfaces}.
\newblock {\em AIAA J.}, 50:1281, 2012.

\bibitem{Weder_Thesis_2012}
M~Weder.
\newblock Linear stability and acoustics of a subsonic plane jet flow.
\newblock M.Sc. Thesis, 2012.

\end{thebibliography}
